\documentclass[pra,twocolumn,tightenlines,showpacs,nofootinbib]{revtex4}
\usepackage{bm,dcolumn,amsmath,graphicx,amsfonts,amssymb}

\begin{document}
\title{Parity nonconservation in hyperfine transitions}

\author{V. A. Dzuba}
\affiliation{School of Physics, University of New South Wales,
Sydney, NSW 2052, Australia}
\author{V. V. Flambaum}
\affiliation{School of Physics, University of New South Wales,
Sydney, NSW 2052, Australia}

\date{ \today }

\begin{abstract}
We use relativistic Hartree-Fock and correlation potential methods to
calculate nuclear spin-dependent parity non-conserving amplitudes
(dominated by the nuclear anapole moment)
between hyperfine structure components of the ground state of odd
isotopes of K, Rb, Cs, Ba$^+$, Yb$^+$, Tl, Fr, and Ra$^+$.
The results are 
to be used for interpretation of current and future measurements.
\end{abstract}
\maketitle

\section{Introduction}

Current best atomic test of the standard model comes from the
measurements~\cite{Wood} and interpretation~\cite{Cs-cor}
of the parity-nonconservation (PNC) in atomic cesium (see also review
\cite{Ginges}). The PNC effect measured in cesium is dominated by
the nuclear spin-independent contribution due to the weak nuclear charge
$Q_W$. The value of $Q_W(^{133}{\rm Cs})$ extracted from the
measurements is in perfect agreement with the standard model.

Since it is hard to compete with cesium PNC in terms of the accuracy
of the interpretation of the measurements, the focus of the atomic PNC
studies has shifted mostly to the study of the nuclear spin-dependent
PNC effects (dominated by the nuclear P-odd anapole moment) and PNC in the
chain of isotopes. The anapole moments has 
been measured so far only for the nucleus of $^{133}$Cs~\cite{Wood}. The
information about weak nuclear forces extracted from these
measurements seems to be  inconsistent with the information obtained from
the limit on the nuclear anapole extracted from the 
thallium PNC~\cite{Vetter,Kozlov02} and some
 other sources (see, e.g.~\cite{Ginges}). 
  Therefore, it would be very important to get more anapole measurements. 
  The experiments are
in  progress or planned for Yb and Dy at Berkeley~\cite{BudkerYbPNC,DyPNC},
Yb$^+$ at Los Alamos~\cite{Torgerson}, Ra$^+$ at
KVI~\cite{Wansbeek,Versolato}, Rb and Fr at
TRIUMF~\cite{SOG10,GOS06}. There were also proposals to measure
atomic PNC in K~\cite{K-AM}, Ba$^+$~\cite{F93}, and
Xe~\cite{BR11}. Most of these experiments are aimed at both, nuclear
spin-dependent PNC effects and PNC in a chain of isotopes. 
The proposals for K~\cite{K-AM}, Rb and
Fr~\cite{SOG10,GOS06} specificly target the effect of the 
the nuclear anapole moment in atomic hyperfine transitions.
 The interpretation of these measurements requires atomic calculations. 
Note that PNC in the Zeeman transitions considered in Ref. \cite{Angstmann}
is also based on such calculations.

Nuclear spin-dependent PNC effects in hyperfine transitions were
considered semiempirically in Refs.~\cite{Novikov,Gorshkov}.
{\em Ab initio} calculations for Fr,
which included correlations, core polarization and Breit interaction
were reported in Ref.~\cite{Porsev01}. Ref.~\cite{JSS03} presents
calculations of the nuclear spin-dependent PNC for a wide range of
single-valence-electron atoms. The calculations are in random phase
approximation (RPA) which means that core polarization is included.

In present paper we perform calculations of the nuclear spin-dependent
PNC for a range of atoms of potential experimental interest. These
includes K, Rb, Cs, Fr, Tl, Ba+, Ra+, and Yb+. We have included the core
polarization (RPA) and correlation effects.
The results in the RPA approximation are in perfect agreement with
Ref.~\cite{JSS03}. 
The inclusion of the
correlations  in our work  increases the PNC amplitudes by 3 to 7\%.
The relatively small magnitude of the correlation corrections is due
to cancelation of different contributions. Based on the detailed study
of the error budget in Ref. \cite{Cs-cor}, we  expect that the
accuracy of present calculations is few per cent.
The agreement with  Ref.~\cite{Porsev01} for Fr is not so good,  and
we discuss the reasons for the difference.

\section{Theory}

\label{theory}
Hamiltonian describing the nuclear spin dependent parity-nonconserving
electron-nuclear interaction can be written in a form
(we use atomic units: $\hbar = |e| = m_e = 1$):
\begin{equation}
     H_{\rm SD-PNC} 
      = \frac{G_F}{\sqrt{2}}                             
     \frac{\varkappa}{I}
     {\bm \alpha} {\bm I} \rho({\bm r}),
\label{eq:H1}
\end{equation}
where  $G_F \approx 2.2225 \times 10^{-14}$ a.u. is the Fermi constant of
the weak interaction, 
$\bm\alpha=\left(
\begin{array}
[c]{cc}%
0 & \bm\sigma\\
\bm\sigma & 0
\end{array}
\right)$ is the Dirac matric, $\bm I$ is the
nuclear spin, and $\rho({\bf r})$ is the nuclear density normalized to 1.
The strength of the spin-dependent PNC interaction is proportional to
the dimensionless constant $\varkappa$ which is to be found from the
measurements. 
There are three major contributions to
$\varkappa$ arising from (i) electromagnetic interaction of atomic
electrons with nuclear {\em anapole moment} ~\cite{FKh80}, (ii)
electron-nucleus spin-dependent weak
interaction~\cite{Novikov,Novikov1} , and (iii) combined effect of the  
spin-independent weak interaction and the magnetic hyperfine
interaction~\cite{FKh85}
(see also review~\cite{Ginges}). In this work we do not distinguish
between different contributions to $\varkappa$ and present the results
in terms of total $\varkappa$ which is the sum of all possible
contributions
 (the nuclear anapole gives the dominating contribution in heavy atoms). 

The PNC amplitude of an electric dipole transition between states of
the same parity $|i\rangle$ and $|f \rangle$ is equal to:
\begin{eqnarray}
   E1^{PNC}_{fi}  &=&  \sum_{n} \left[
\frac{\langle f | {\bm d} | n  \rangle
      \langle n | H_{\rm PNC} | i \rangle}{E_i - E_n}\right.
\nonumber \\
      &+&
\left.\frac{\langle f | H_{\rm PNC} | n  \rangle
      \langle n | d_q | i \rangle}{E_f - E_n} \right],
\label{eq:e2}
\end{eqnarray}
where ${\bm d} = -e\sum_i {\bm r_i}$ is the electric dipole operator,
  $|a \rangle \equiv |J_a F_a M_a \rangle$ and ${\bm F} = {\bm I}
+ {\bm J}$ is the total angular momentum. 

Applying the Wigner-Eckart theorem we can express the amplitudes via
reduced matrix elements
\begin{eqnarray}
  E1^{PNC}_{fi} &=&
      (-1)^{F_f-M_f} \left( \begin{array}{ccc}
                           F_f & 1 & F_i  \\
                          -M_f & q & M_i   \\
                           \end{array} \right) \nonumber \\
   &\times& \langle J_f F_f || d_{\rm PNC} || J_i F_i \rangle .
\end{eqnarray}
Detailed expressions for the reduced matrix elements of the SI and
SD PNC amplitudes can be found e.g. in Refs.~\cite{Porsev01} and
\cite{JSS03}.
For the SD PNC amplitude we have
\begin{eqnarray}
    && \langle J_f,F_f || d_{\rm SD} || J_i,F_i \rangle =
    \frac{G_F}{\sqrt{2}} \varkappa \nonumber \\
     &&\times  \sqrt{(I+1)(2I+1)(2F_i+1)(2F_f+1)/I}  \nonumber \\
    &&\times
     \sum_{n} \left[ (-1)^{J_f - J_i}
     \left\{ \begin{array}{ccc}
     J_n  &  J_i  &   1    \\
      I   &   I   &  F_i   \\                                  
     \end{array} \right\}
     \left\{ \begin{array}{ccc}
      J_n  &  J_f  &  1   \\
      F_f  &  F_i  &  I   \\
     \end{array} \right\} \right. \nonumber \\
  &&\times \frac{ \langle J_f || {\bm d} || n, J_n \rangle
     \langle n, J_n || {\bm \alpha}\rho || J_i \rangle }{E_n -
     E_i} \label{eq:dsd}  \\
  &&+
     (-1)^{F_f - F_i}
     \left\{ \begin{array}{ccc}
     J_n  &  J_f  &   1    \\
      I   &   I   &  F_f   \\
     \end{array} \right\}
     \left\{ \begin{array}{ccc}
     J_n  &  J_i  &  1   \\
     F_i  &  F_f  &  I   \\
     \end{array} \right\} \nonumber \\
 &&\times
     \left. \frac{\langle J_f || {\bm \alpha}\rho ||n,J_n \rangle
            \langle n,J_n || {\bm d} ||J_i \rangle}{E_n - E_f}  \right].
\nonumber
\end{eqnarray}
This formula can be used for optical and microwave transitions. For
the microwave transitions it can be further simplified as it has been
done in Ref.~\cite{Porsev01}.

To calculate the PNC amplitudes we use direct summation over a
complete set of single-electron states constructed using the B-spline
technique~\cite{Bspline}. Correlations and core polarization effects
are included by means of the correlation potential
method~\cite{CPM}. The calculations are very similar to those of our
previous work~\cite{BaRa}, therefore we omit here the details.

\section{Results}

\begin{table*}
\caption{Reduced matrix elements of the nuclear spin-dependent weak interaction
for microwave transitions between ground-state hyperfine levels
$F_1,F_2$ in atoms and ions of potential experimental interest. In
this table, $Z$ is the nuclear charge, $A$ is the atomic number, $I$
is the nuclear spin, the 
hyperfine levels have angular momentum $F=I \pm 1/2$, $M_1$ is
magnetic dipole transition amplitude, $P$ is the degree of circular
polarization. All dimensional numbers are in atomic units. The values
for $M1$ amplitudes include Bohr magneton ($\mu_B=\alpha/2$ in atomic units).
The PNC amplitudes and circular polarization $P$ are proportional to the
weak interaction constant $\xi$ (\ref{eq:H2}) which is omitted from
the table. 
Numbers in square brackets represent powers of 10.} 
\label{t:1}
\begin{ruledtabular}
\begin{tabular}{lcrccc ccc lc}
Atom & $Z$ & \multicolumn{1}{c}{$A$} & $I$ & $F_1$ & $F_2$ & RPA & RPA &
RPA+$\Sigma$ & \multicolumn{1}{c}{$M_1$} & $P$ \\
   &     &  &   &   &   & Ref.~\cite{JSS03} & \multicolumn{4}{c}{this
     work} \\
\hline
K     & 19 & 39 & 3/2 & 2 & 1 & 2.222[-13] & 2.222[-13] & 2.378[-13] & 0.0100  & 4.76[-11] \\
Rb    & 37 & 85 & 5/2 & 3 & 2 & 2.550[-12] & 2.532[-12] & 2.660[-12] & 0.0125 & 4.27[-10]\\
Rb    & 37 & 87 & 3/2 & 2 & 1 & 1.363[-12] & 1.353[-12] & 1.422[-12] & 0.0100 & 2.85[-10]\\
Cs    & 55 &133 & 7/2 & 4 & 3 & 1.724[-11] & 1.720[-11] & 1.819[-11] & 0.0145 & 2.51[-09]\\
Ba$^+$& 56 &135 & 3/2 & 2 & 1 & 6.169[-12] & 6.186[-12] & 6.385[-12] & 0.0100   & 1.28[-09]\\
Yb$^+$& 70 &171 & 1/2 & 1 & 0 &            & 3.844[-12] & 3.987[-12] & 0.0063 & 1.26[-09]\\
Yb$^+$& 70 &173 & 5/2 & 3 & 2 &            & 2.274[-11] & 2.359[-11] & 0.0125 & 3.79[-09]\\
Tl    & 81 &203 & 1/2 & 1 & 0 & 3.000[-11] & 3.001[-11] & 3.066[-11] & 0.0021 & 2.92[-08]\\
Fr    & 87 &211 & 9/2 & 5 & 4 & 2.379[-10] & 2.362[-10] & 2.487[-10] & 0.0162 & 3.07[-08]\\
Fr    & 87 &223 & 3/2 & 2 & 1 & 5.820[-11] & 5.815[-11] & 6.051[-11] & 0.0100   & 1.21[-08]\\
Ra$^+$& 88 &223 & 3/2 & 2 & 1 & 5.987[-11] & 5.991[-11] & 6.147[-11] & 0.0100   & 1.23[-08]\\
\end{tabular}
\end{ruledtabular}
\end{table*}

The results of calculations are presented in Table~\ref{t:1}. The
results of present work for the PNC amplitude are presented in two
different approximations, the RPA approximation, which includes core
polarization but no correlations beyond it, and final results which
include core polarization and Brueckner-type correlations. The results
of Ref.~\cite{JSS03}, which were obtained in the RPA approximation,
are also presented for comparison. For the convenience of the
comparison we present the results for the PNC amplitudes in a form
which corresponds to the PNC Hamiltonian, used in \cite{JSS03}:
\begin{equation}
     H_{\rm SD-PNC} = \frac{G_F}{\sqrt{2}}\xi
     {\bm \alpha} {\bm I} \rho({\bm r}).
\label{eq:H2}
\end{equation}
It differs from (\ref{eq:H1}) by the use of the different weak
interaction constant $\xi$ ($\xi = \varkappa/I$).

The RPA results of \cite{JSS03} and present work are in perfect
agreement with each other. The inclusion of Brueckner-type
correlations increase the PNC amplitudes by 3 to 7\%. It is
interesting to note that the correlation correction is larger for
light elements and slowly decrease for higher $Z$. 

In Table \ref{t:1} we also present the values of the reduced matrix
elements for the magnetic dipole ($M1$) transition amplitudes between
corresponding hyperfine states and the degrees of circular
polarization of light $P$. The $M1$ amplitudes are given by
\begin{eqnarray}
 &&\langle F,J||M1||F^{\prime},J \rangle =
 (-1)^{I+J+F}g(J,L) \mu_B \label{eq:m1g} \\ 
&& \times \sqrt{(2F+1)(2F^{\prime}+1)J(J+1)(2J+1)}
\left\{ \begin{array}{ccc} F & 1 & F^{\prime} \\ J & I & 
J \end{array} \right\},\nonumber
\end{eqnarray}
where $g(J,L)$ is the $g$-factor of atomic state $J,L,S=1/2$
\begin{equation}
  g(J,L) = 1 + \frac{J(J+1)-L(L+1)+\frac{3}{4}}{2J(J+1)},
\label{eq:g}
\end{equation}
$J,L$ are atomic total and angular momentums, and $\mu_B =
|e|\hbar/2mc$ is Bohr magneton. In the case of $J=1/2, F=I \pm 1/2$
the Eq. (\ref{eq:m1g}) can be further reduced to
\begin{equation}
  M1 =
  2\sqrt{\frac{I(I+1)}{I+\frac{1}{2}}}\left[1-\frac{L(L+1)}{3}\right]\mu_B.
\label{eq:m1}
\end{equation}
The degree of the circular polarization of light is given by 
\begin{equation}
 P = 2 \frac{Im(E1)}{M1}. 
\label{eq:p}
\end{equation}

The results for $^{211}$Fr, converted to a different definition of
the weak interaction constant ($\varkappa=\xi I$) consistent with
Hamiltonian (\ref{eq:H1}), are $0.529 
\times 10^{-10} i\varkappa$~\cite{JSS03} and $0.553\times 10^{-10}
i\varkappa$ (present work). They differ by only 4.5\% and this
difference is due to correlations which were included in the present work
but not included in Ref.~\cite{JSS03}. On the other hand, the
difference between the result of present work and the calculations by
Porsev and Kozlov~\cite{Porsev01}, which is $0.491\times 10^{-10}
i\varkappa$, is significantly larger, being about 13\%. It is important
to understand the reason for this difference, since the calculations by
Porsev and Kozlov are the only other calculations for Fr which
included correlations beyond the RPA approximation.
The experimental work  for Fr is in progress at
TRIUMF~\cite{SOG10,GOS06} and for its future interpretation it is
important to have reliable theoretical results.

\begin{table}
\caption{Contributions to the $6s-7s \ E_{\rm PNC}$ for Cs in units
  $10^{-11}iea_B(-Q_W/N)$. First line is equivalent to the
  approximation used in present work.}
\label{t:Cs}
\begin{ruledtabular}
\begin{tabular}{lr}
Brueckner orbitals + core polarization &  0.9077 \\
Weak correlation potential &  0.0038 \\
Structural radiation       &  0.0029 \\
Normalization              & -0.0066 \\
Breit correction           & -0.0055 \\
Neutron distribution       & -0.0018 \\
Radiative corrections\footnotemark[1] & -0.0029 \\
Total                      &  0.8976 \\
\end{tabular}
\footnotetext[1]{Ref.~\cite{FG05}}
\end{ruledtabular}
\end{table}

The most obvious difference between present calculations and those of
Ref.~\cite{Porsev01} is inclusion of higher-order correlations in our
work and some small effects in~\cite{Porsev01}.
These small effects include Breit interaction, structural radiation and
renormalization of wave functions. We first discuss these
small corrections. The most detailed study of all important
corrections to the PNC amplitude has been done for the $6s-7s$ PNC
amplitude for Cs~\cite{Cs-cor,FG05}. The results are summarized in
Table~\ref{t:Cs}. The relative values of specific corrections for the
hfs PNC transition in Fr might be slightly different, however, the
qualitative picture should be very similar. First line in
Table~\ref{t:Cs} corresponds to the approximation used in the present
work. Note that the final result is only about 1\% smaller. Furthermore,
there is strong cancelation between different contributions. For
example, the contributions from the weak correlation potential and
the structural radiation are canceled by the renormalization of the wave
functions. 
Different contributions to the radiative corrections are not so small
(up to -0.8\%). However, they have different signs and nearly cancel
each other.
This means that inclusion of some of the small corrections
while not including others cannot be justified and may lead to less
accurate results.

\begin{table}
\caption{Reduced matrix elements of the nuclear spin-dependent PNC
  amplitude $\langle 7s,F=5|| d_{\rm PNC}||7s,F=4\rangle$ for
  $^{211}$Fr in units $\times 10^{-10}i\varkappa$. To compare these
  numbers with those presented in Table \ref{t:1} one should multiply
  them by the nuclear spin $I=9/2$. Contributions from the core are not
included in first three columns.}
\label{t:2}
\begin{ruledtabular}
\begin{tabular}{lrrcrr}
 & \multicolumn{1}{c}{DHF} &\multicolumn{1}{c}{RPA}
 &\multicolumn{1}{c}{MBPT\footnotemark[1]} & \multicolumn{1}{c}{core}
 &\multicolumn{1}{c}{Total} \\
\hline
Porsev and Kozlov~\cite{Porsev01} & 0.418 & 0.058 & 0.033 & -0.018 & 0.491 \\
this work                     & 0.423 & 0.058 & 0.009 &  0.063 & 0.553 \\
\end{tabular}
\footnotetext[1]{Many-body perturbation theory for the correlation correction.}
\end{ruledtabular}
\end{table}

We proceed to the examining the correlations.
For this purpose we compare our results with those from
Ref.~\cite{Porsev01} term by term as it is shown in
Table~\ref{t:2}. The first column shows the PNC amplitude in the
Dirac-Hartree-Fock approximation without any correlations. There is
a small, about 1\% difference in the results. Some of this difference
might be attributed to the Breit interaction which is included in
\cite{Porsev01} and not included in our work. There might be also some
difference due to different treatment of the nucleus. We use smooth
Fermi distribution of the nuclear electric charge and anapole moment,
while a step-like function is used in \cite{Porsev01}.

The RPA correction is the same in both works (see Table \ref{t:2}).
The difference in correlation correction
is significant which is not surprising due to very different treatment
of the correlations in two works. We include the dominating Brueckner-type
correlations with the use of the all-order correlation potential
$\Sigma$~\cite{BaRa}. The correlations are included in
Ref.~\cite{Porsev01} in the second-order only, including
the structure radiation and the renormalization of the wave functions. 
These latter corrections are small (see discussion above) and cannot
explain the difference in the results. 

To test whether the difference is due to the higher-order correlations
we performed calculations in which all higher-order correlations were
removed.  The correlation potential
$\Sigma$ was calculated in the second order and the core polarization
corrections were not included. The result,
$0.021 \times 10^{-10}i\varkappa$, is in a reasonable agreement
with~\cite{Porsev01}. 
In the end, the difference in the results due to different treatment of
the correlations is about 5\%. This difference is most likely due to
the higher-order correlations included in the present work. 

The largest difference comes from the contribution of the core
states. This contribution is small and negative in
Ref.~\cite{Porsev01}. In our calculations it is not so small and it is
positive (see Table \ref{t:2}). The value of this correction in
\cite{Porsev01} suggests that RPA corrections were probably not
included. Our value for the core contributions without RPA corrections
is $+0.015 \times 10^{-10}i\varkappa$. However, there is no clear explanation
for the different sign. Note, that we do not distinguish between core
and excited states in the summation over complete set of states in
(\ref{eq:dsd}). This leaves no room for a sign error. On the other
hand, to the best of our knowledge, the contribution of the core
states was calculated separately in Ref.~\cite{Porsev01}. 

There is a simple test to check the sign of the core contribution. The
summation over core states in (\ref{eq:dsd}) is dominated by the
$6p_{1/2}$ state, while summation above core is dominated by the
$7p_{1/2}$ state. The angular coefficients in (\ref{eq:dsd}) are the
same for core and higher states. One needs only to compare the energy
denominators and the matrix elements of weak and electric dipole
interactions. The energy denominators $E_{7s}-E_{6p_{1/2}}$  and
$E_{7s}-E_{7p_{1/2}}$ are of the opposite sign. To compare the signs of
matrix elements we need to fix the phase of the wave functions. It is
convenient to have $f(r)>0$ at $r \rightarrow 0$, where $f(r)$ is the
upper component of the Dirac spinor. Then, the $\langle
7s||H_{\rm PNC}||6p_{1/2} \rangle$ and $\langle 7s||H_{\rm
  PNC}||7p_{1/2} \rangle$   matrix elements have the same sign since
the values of these matrix elements comes from short distances where
the $6p_{1/2}$ and $7p_{1/2}$ functions are proportional to each
other~\cite{ratios}. In contrast, the signs of the $\langle
7s||d||6p_{1/2} \rangle$ and $\langle 7s||d||6p_{1/2} \rangle$ matrix
elements are different since their value comes from large distances
where the $6p_{1/2}$ and $7p_{1/2}$ functions have different sign.
This statement can be checked using the simplest possible approximation,
e.g. the DHF approximation. The many body corrections are not large
enough to change the sign of the electric dipole matrix elements.

In our previous work~\cite{ratios}, we used the results of
Ref.~\cite{Porsev01} and the ratio of the matrix elements of
spin-dependent PNC interaction and electron electric dipole moment
(EDM) to extract the value of the EDM enhancement factor for Fr from
the spin-dependent PNC calculations of \cite{Porsev01}. The result,
$d({\rm Fr}) = 854 d_e$, was in reasonably good agreement with the
many-body calculations of Ref.~\cite{BDFM}, $d({\rm Fr}) = 910(46)
d_e$. We can use the result of present work instead of \cite{Porsev01}
for the same purpose. First, we need to remove the contribution of the
$p_{3/2}$ states to use the proportionality of the matrix
elements. This reduces the PNC amplitude for $^{211}$Fr to $0.523
\times 10^{-10}\varkappa$.  The
resulting EDM enhancement factor $d({\rm Fr}) = 911$ is even in better
agreement with the result 910(46) of Ref.~\cite{BDFM}. This is a good
consistency test of the calculations.

\section*{Acknowledgments}

The authors are grateful to M. G. Kozlov for useful discussion.
The work was supported in part by the Australian Research Council.


\begin{thebibliography}{99} 

\bibitem{Wood} C. S. Wood, S. C. Bennett, D. Cho, B. P. Masterson,
  J. L. Roberst, C. E. Tanner, C. E. Wieman, Science {\bf 275}, 1759
  (1997).

\bibitem{Cs-cor}  V. A. Dzuba, V. V. Flambaum, and J. S. M. Ginges,  Phys.
Rev. D {\bf 66}, 076013 (2002);
S. G. Porsev, K. Beloy, and A. Derevianko, 
Phys. Rev. Lett. {\bf 102}, 181601 (2009);
S. G. Porsev, K. Beloy, and A. Derevianko,
Phys. Rev. D {\bf 82}, 036008 (2010).

\bibitem{Ginges} J. S. M. Ginges and V. V. Flambaum,
Phys. Rep. {\bf 397}, 63 (2004).

\bibitem{Vetter} P. A. Vetter, D. M. Meekhof, P. K. Majumder, 
  S. K. Lamoreaux, and E. N. Fortson,
  Phys. Ref. Lett. {\bf 74}, 2658 (1995).

\bibitem{Kozlov02} M. G. Kozlov, Pis'ma Zh. Eksp. Teor. Fiz. {\bf 75},
  651 (2002) [Sov. Phys. JETP Lett. {\bf 75}, 534 (2002)].

\bibitem{BudkerYbPNC} K. Tsigutkin, D. Dounas-Frazer, A. Family,
  J. E. Stalnaker, V. V. Yashchuk, and D. Budker, 
  Phys. Rev. Lett. {\bf 103}, 071601 (2009); 
  Phys. Rev. A {\bf 81}, 032114 (2010).

\bibitem{DyPNC}  A. T. Nguyen, D. Budker, D. DeMille, and M. Zolotorev,
Phys. Rev. A \textbf{56}, 3453 (1997).

\bibitem{Torgerson} J. Torgerson, private communication (2010).

\bibitem{Wansbeek} L. W. Wansbeek {\em et al}, Phys. Rev. A {\bf 78},
  050501(R) (2008).

\bibitem{Versolato} O. O. Versolato {\em et al}, Phys. Rev. A {\bf 82},
  010501(R) (2010).

\bibitem{SOG10} D. Sheng, L. A. Orozco, and E. Gomez, 
J. Phys. B {\bf 43}, 074004 (2010).

\bibitem{GOS06} E. Gomez, L. A. Orozco, and G. D. Sprouse, 
Rep. Prog. Phys. {\bf 69}, 79 (2006).

\bibitem{Angstmann} E. J. Angstmann, T. H. Dinh, and V. V.
Flambaum,  Phys. Rev. A {\bf72}, 052108 (2005).

\bibitem{K-AM} V. F. Ezhov, M. G. Kozlov, G. B. Krygin GB, {\em et  
    al}, Tech. Phys. Lett. {\bf 30}, 917 (2004).   

\bibitem{F93} N. Fortson, Phys. Rev. Lett. {\bf 70}, 2383 (1993).

\bibitem{BR11} D. Budker and T. P. Rakitzis, private communication (2011).

\bibitem{Novikov} V. N. Novikov and I. B. Khriplovich, Pis'ma
  Zh. Eksp. Teor. Fiz. {\bf 22}, 162 (1975) [JETP Lett. {\bf 22}, 74 (1975)].

\bibitem{Gorshkov} V. G. Gorshkov, V. F. Ezhov, M. G. Kozlov, and
A. I. Mikhailov, Yad. Fiz. {\bf 48}, 1363 (1988)
[Sov. J. Nucl. Phys. {\bf 48}, 867 (1988)].

\bibitem{Porsev01} S. G. Porsev and M. G. Kozlov,
Phys. Rev. A {\bf 64}, 064101 (2001). 

\bibitem{JSS03} W. R. Johnson, M. S. Safronova, and U. I. Safronova,
Phys. Rev. A {\bf 67}, 062106 (2003).



\bibitem{FKh80} V. V. Flambaum, I. B. Khriplovich, 
  ZhETP {\bf 79}, 1656 (1980) (Soviet Phys. JETP {\bf 52}, 835 (1980)).
  V.V. Flambaum, I.B. Khriplovich, O.P. Sushkov.
Phys. Lett. B. {\bf 146}, 367, 1984.

\bibitem{Novikov1} V. N. Novikov, O. P. Sushkov, V. V. Flambaum,
  I. B. Khriplovich, 
  ZhETP {\bf 73}, 802 (1977) (Soviet Phys. JETP {\bf 46}, 420 (1977)).

\bibitem{FKh85} V. V. Flambaum, I. B. Khriplovich, 
  ZhETP {\bf 89}, 1505 (1985) (Soviet Phys. JETP {\bf 62}, 872 (1985)).

\bibitem{Bspline} W. R. Johnson, and J. Sapirstein, 
      Phys. Rev. Lett. {\bf 57}, 1126 (1986).

\bibitem{CPM} V. A. Dzuba, V. V. Flambaum, P. G. Silvestrov, O. P. Sushkov,
J. Phys. B {\bf 20}, 1399 (1987).

\bibitem{BaRa} V. A. Dzuba and V. V. Flambaum,                            
      Phys. Rev. A {\bf 83}, 052513 (2011).

\bibitem{FG05} V. V. Flambaum and J. S. M. Ginges,
Phys. Rev. A {\bf 72}, 052115 (2005).

\bibitem{ratios} V. A. Dzuba, V. V. Flambaum, and C. Harabati,
      Phys. Rev. A {\bf 84}, 052108 (2011).

\bibitem{BDFM} T. M. R. Byrnes, V. A. Dzuba, V. V. Flambaum, and D. W. Murray,
Phys. Rev. A, {\bf 59}, 3082 (1999).


\end{thebibliography}
\end{document}